# Parsimonious Adaptive Rejection Sampling

Luca Martino*

* Image Processing Laboratory, Universitat de València (Spain).

**Abstract**

Monte Carlo (MC) methods have become very popular in signal processing during the past decades. The adaptive rejection sampling (ARS) algorithms are well-known MC technique which draw efficiently independent samples from univariate target densities. The ARS schemes yield a sequence of proposal functions that converge toward the target, so that the probability of accepting a sample approaches one. However, sampling from the proposal pdf becomes more computationally demanding each time it is updated. We propose the Parsimonious Adaptive Rejection Sampling (PARS) method, where an efficient trade-off between acceptance rate and proposal complexity is obtained. Thus, the resulting algorithm is faster than the standard ARS approach.

## 1 Introduction

Adaptive rejection sampling (ARS) schemes are widely employed in signal processing for optimization, complex system simulation and Bayesian inference [8, 9, 10, 12]. They generate independent samples from a target probability density function (pdf). For instance, the ARS algorithms are required within Gibbs-type samplers for drawing one (or several; see [5]) samples from the univariate full-conditional pdfs [2, 9, 5]. Since the standard ARS method [2] can be applied only when the target density is log-concave, several extensions have been proposed [1, 3, 4, 7, 8, 6, 9].

In this letter, we focus on the computational cost required by ARS. The ARS algorithms obtain high acceptance rates building a sequence of non-parametric proposal functions which become closer and closer to target function. Hence, this improvement of the acceptance rate is obtained building more complex proposals, i.e., more computationally demanding. The overall time spent by an ARS scheme depends on (a) the acceptance rate and (b) the time required for drawing from the proposal pdf. In ARS, a trade-off is found decreasing the probability of updating the proposal as the acceptance rate grows. Here, we introduce the *Parsimonious Adaptive Rejection Sampling* (PARS) method, which achieves a better compromise between acceptance rate and proposal complexity. PARS obtains a better construction of the non-parametric proposal pdf, reaching high acceptance rates with a smaller complexity of the proposal w.r.t. to the classical ARS approach. As a consequence, PARS is faster than ARS as confirmed by the numerical

simulations.The Matlab code of PARS and ARS, related to the provided numerical results, is given at `Matlab-File Exchange` webpage.

## 2   Standard ARS

Let us denote the target density as $\bar{\pi}(x) \propto \pi(x) = \exp(V(x))$, $x \in \mathcal{X} \subseteq \mathbb{R}$. The adaptive proposal pdf is denoted as $\bar{q}_t(x|\mathcal{S}_t) \propto q_t(x|\mathcal{S}_t) = \exp(W_t(x))$, where $t \in \mathbb{N}$. In order to apply rejection sampling (RS) [8, 12], it is necessary to build $q_t(x|\mathcal{S}_t)$ as an envelope function of $\pi(x)$, i.e.,

$$q_t(x|\mathcal{S}_t) \geq \pi(x), \quad \text{or} \quad W_t(x) \geq V(x), \qquad (1)$$

for all $x \in \mathcal{X}$ and $t \in \mathbb{N}$. Let us assume that $V(x) = \log \pi(x)$ is concave (i.e., $\pi(x)$ is log-concave), and we are able to evaluate the function $V(x)$ and its first derivative $V'(x)$. The standard ARS technique [2] considers a set of support points (nodes) at the $t$-th iteration, $\mathcal{S}_t = \{s_1, s_2, \ldots, s_{m_t}\} \subset \mathcal{X}$, with $s_1 < \ldots < s_{m_t}$ and $m_t = |\mathcal{S}_t|$, in order to construct a non-parametric envelope function $q_t(x|\mathcal{S}_t)$. We denote as $w_i(x)$ as the straight line tangent to $V(x)$ at $s_i$ for $i = 1, \ldots, m_t$. We can build a piecewise linear function as

$$W_t(x) = \min[w_1(x), \ldots, w_{m_t}(x)], \quad x \in \mathcal{X}. \qquad (2)$$

Hence, the proposal function defined as $q_t(x|\mathcal{S}_t) = \exp(W_t(x))$ is formed by exponential pieces, where $W_t(x) \geq V(x)$ so that $q_t(x|\mathcal{S}_t) \geq \pi(x)$. Figure 1 depicts an example of piecewise linear function $W_t(x)$ built with $m_t = 3$ support points. Several other construction procedures for specific non-log-concave targets $\pi(x)$ have been proposed [4, 8]. Table 1 summarizes the ARS algorithm for drawing $N$ independent samples from $\bar{\pi}(x)$. At each iteration $t$, a sample $x'$ is drawn from $\bar{q}_t(x|\mathcal{S}_t)$ and accepted with probability $\frac{\pi(x')}{q_t(x'|\mathcal{S}_t)}$. Note that a new point is added to the support set $\mathcal{S}_t$ whenever $x'$ is rejected in the RS test, so that $q_{t+1}$ becomes closer to $\pi$. Denoting as $T$ the total number of iterations of the algorithm, we have always $T \geq N$ owing to the $T - N$ rejected samples.

## 3   Computational cost of ARS

The computational cost of an ARS-type method depends on two elements:

1. The acceptance rate at $t$-th iteration,

$$\eta_t = \int_{\mathcal{X}} \frac{\pi(x)}{q_t(x|\mathcal{S}_t)} \bar{q}_t(x|\mathcal{S}_t) dx, \qquad (3)$$

    where $0 \leq \eta_t \leq 1$, $\forall t \in \mathbb{N}$, by construction. As $q_t(x|\mathcal{S}_t)$ becomes closer to $\pi(x)$ for $t \to \infty$ then $\eta_t \to 1$. In the ideal case $\eta_t = 1$ for all $t$, hence $T = N$.

2. The computational time required for sampling from $\bar{q}_t(x|\mathcal{S}_t)$.

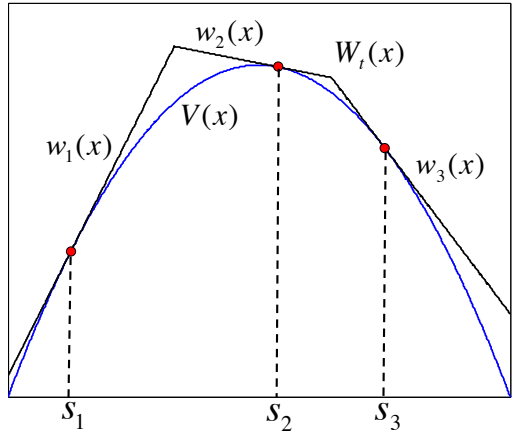

**Figure 1:** Example of construction of the piecewise linear function $W_t(x)$ with $m_t = 3$ support points, such that $W_t(x) \geq V(x)$.

**Table 1:** Standard Adaptive Rejection Sampling

---

1. Set $t = 0$ and $n = 0$. Choose an initial set $\mathcal{S}_0 = \{s_1, \ldots, s_{m_0}\}$.

2. **While $n < N$:**

    (a) Build the proposal $q_t(x|\mathcal{S}_t)$ according to Eq. (2).

    (b) Draw $x' \sim \bar{q}_t(x|\mathcal{S}_t)$ and $u' \sim \mathcal{U}([0,1])$.

    (c) If $u' \leq \frac{\pi(x')}{q_t(x'|\mathcal{S}_t)}$, then set $x_{n+1} = x'$ and $\mathcal{S}_{t+1} = \mathcal{S}_t$.

    (d) Otherwise, if $u' > \frac{\pi(x')}{q_t(x'|\mathcal{S}_t)}$, update
    $$\mathcal{S}_{t+1} = \mathcal{S}_t \cup \{x'\}.$$

    (e) Set $t = t + 1$ and $n = n + 1$.

3. **Outputs:** The $N$ accepted samples, $x_1, \ldots, x_N$.

---

We desire that the acceptance rate is close to 1 and, simultaneously, that the computational time required for drawing from $q_t(x|\mathcal{S}_t)$ is small. Note that an increase of the acceptance rate requires the use of a more complicated proposal density $q_t(x|\mathcal{S}_t)$. ARS provides a possible compromise choosing the support points adequately by the adaptation. Indeed, let us define the $L_1$ distance between $q_t$ and $\pi$ as

$$D(q_t, \pi) = \|q_t(x|\mathcal{S}_t) - \pi(x)\|_1 = \int_{\mathcal{X}} |q_t(x|\mathcal{S}_t) - \pi(x)| dx. \tag{4}$$

ARS ensures that $D(q_t, \pi) \to 0$ when $t \to \infty$ (and $\eta_t \to 1$) since $q_t$ becomes closer to $\pi$ as more nodes are included (see Figure 1). The probability of adding a new support point

$$P_t = 1 - \eta_t = \frac{D(q_t, \pi)}{\int_{\mathcal{X}} q_t(x|\mathcal{S}_t)dx}, \qquad (5)$$

tends to zero as $t \to \infty$, since $D(q_t, \pi) \to 0$. Hence, the number of nodes $m_t$ tends to saturate as $t \to \infty$ (i.e., $m_t$ as function of $t$ is increasing but convex).

Let us denote the exponential pieces as $h_n(x) = \exp(w_n(x))$, $n = 1, \ldots, N$, so that

$$q_t(x|\mathcal{S}_t) = h_n(x), \quad \text{for} \quad x \in \mathcal{I}_n = (e_{n-1}, e_n), \quad n = 1, \ldots, N,$$

where $e_n$ is the intersection point between the straight lines $w_n(x)$ and $w_{n+1}(x)$, for $n = 2, \ldots, N-1$, and $e_0 = -\infty$ and $e_N = +\infty$ (if $\mathcal{X} = \mathbb{R}$). Thus, in order to draw one sample $x'$ from $\bar{q}_t(x|\mathcal{S}_t)$, we need to:

1. Compute analytically the area $A_i = \int_{\mathcal{I}_i} h_i(x)dx$ below each exponential piece, and obtain the normalized weights $\rho_i = \frac{A_i}{\sum_{n=1}^{N} A_n}$.

2. Select an index $j^*$ (i.e., one piece) according to the probability mass $\rho_i$, $i = 1, \ldots, N$.

3. Draw $x'$ from $h_{j^*}(x)$ restricted within the domain $\mathcal{I}_{j^*} = (e_{j^*-1}, e_{j^*})$, and zero outside (i.e., from a truncated exponential pdf).

Note that a multinomial sampling is needed at step 2. Hence, the computational cost of drawing from $\bar{q}_t(x|\mathcal{S}_t)$ increases as the number of points $m_t$ grows.

## 4 Parsimonious ARS

### 4.1 Key observation

In ARS, the probability of adding a new support point $P_t$ vanishes to zero as $t \to \infty$. However, for a finite $t$, we have always a positive probability $P_t > 0$ (although small) of adding a new point, so that a new support point could be incorporated, building a better $q_t(x|\mathcal{S}_t)$ and yielding an increase of the acceptance rate. After a certain iteration $\tau$, i.e., $t > \tau$, this improvement of the acceptance rate could not balance out the increase of the time required for drawing from the proposal, due to the addition of the new point. Namely, if the acceptance rate is enough close to 1, a further addition of a support point could slow down the algorithm, becoming prejudicial.

### 4.2 Novel scheme

In the standard ARS, the addition of a new node is linked to the RS test, indeed all the rejected samples are incorporated as new support points. We propose the *Parsimonious Adaptive Rejection Sampling* (PARS) method, where a different test is considered for adding a new node. Specifically, the RS test used in order to accept or reject the proposed sample $x'$, whereas an additional deterministic test is performed for incorporating (or not) $x'$ in $\mathcal{S}_t$. Given a pre-established threshold $0 \leq \delta \leq 1$, $x'$ is employed as new node if $\frac{\pi(x')}{q_t(x'|\mathcal{S}_t)} \leq \delta$. PARS is outlined in Table 2. If $\delta = 0$, PARS becomes a non-adaptive RS technique whereas, if $\delta = 1$, all the proposed samples $x'$ are added as new nodes. For a generic $0 < \delta < 1$, at some iteration $t^*$ the adaptation is stopped since no more support points are included. PARS forces to obtain acceptance probabilities $\frac{\pi(x')}{q(x'|\mathcal{S}_t)}$ greater than $\delta$, performing a better selection of the nodes. Indeed, in general, PARS obtains high acceptance values using a smaller number of support points than ARS. As a consequence, PARS is faster than ARS as shown in the numerical simulations.

**Table 2:** Parsimonious Adaptive Rejection Sampling (PARS)

1. Set $t = 0$ and $n = 0$. Choose $\mathcal{S}_0 = \{s_1, \ldots, s_{m_0}\}$ and a threshold value $0 \leq \delta \leq 1$.

2. **While $n < N$:**

    (a) Build the proposal $q_t(x|\mathcal{S}_t)$ according to Eq. (2).

    (b) Draw $x' \sim \bar{q}_t(x|\mathcal{S}_t) \propto q_t(x|\mathcal{S}_t)$ and $u' \sim \mathcal{U}([0,1])$.

    (c) If $u' \leq \frac{\pi(x')}{q_t(x'|\mathcal{S}_t)}$, then set $x_{n+1} = x'$.

    (d) If $\frac{\pi(x')}{q_t(x'|\mathcal{S}_t)} \leq \delta$, update
    $$\mathcal{S}_{t+1} = \mathcal{S}_t \cup \{x'\}.$$
    Otherwise, if $\frac{\pi(x')}{q_t(x'|\mathcal{S}_t)} > \delta$, set $\mathcal{S}_{t+1} = \mathcal{S}_t$.

    (e) Set $t = t + 1$ and $n = n + 1$.

3. **Outputs:** The $N$ accepted samples, $x_1, \ldots, x_N$.

# 5 Numerical Results

The Nakagami-$m$ distribution is widely used for the simulation of fading channels in wireless communications, due to its good agreement with empirical channel measurements for some urban multipath environments [11]. The Nakagami pdf is

$$\bar{\pi}(x) \propto \pi(x) = x^{2m-1} \exp\left(-\frac{m}{\Omega}x^2\right), \qquad x \geq 0, \tag{6}$$

where $m \geq 0.5$ is the fading parameter, which indicates the fading depth, and $\Omega > 0$ is the average received power. Several methods for drawing samples from a Nakagami-$m$ pdf have been proposed [10, 12]. In our experiments, we set $m = 1.2$ and $\Omega = 2$. We compare the computational time (computed in a MAC-1.7 GHz-8 GB) required by ARS and PARS (considering $\delta = 0.8$) for drawing $N \in \{5 \cdot 10^4, 10^5, 1.5 \cdot 10^5, 2 \cdot 10^5\}$ samples, with initial set $\mathcal{S}_0 = \{0.5, 1, 2\}$. The results, averaged over 200 runs, are given in Figure 2(a). PARS is always faster than ARS, and the benefit grows with $N$. The corresponding averaged final numbers of nodes, $E[m_T]$, are provided in Figures 2(b). We can observe that the total number of nodes of PARS ($\delta = 0.8$) remains virtually constant. Figures 2(c)-(d) provides the computational time required by PARS and the final number of nodes as function of $\delta$, setting $N = 5 \cdot 10^4$. The values corresponding to ARS are showing with constant dashed lines. We can observe that PARS is always faster than ARS with the exception of the extreme values $\delta = 0.999$ and $\delta = 0.9999$, for which values PARS incorporates higher number of nodes, $E[m_T] = 137.2$ and $E[m_T] = 385.5$, respectively. However, note that these values corresponding to $\delta \in \{0.999, 0.9999\}$, are surprisingly small, since $E[m_T] = N = 5 \cdot 10^4$ for $\delta = 1$, by construction. This is due to the ability of PARS in selecting nodes for building a suitable proposal pdf. Indeed, for $N = 5 \cdot 10^4$, PARS obtains an overall acceptance rate $E\left[\frac{N}{T}\right] = 0.8524$ with only $E[m_T] = 6.75$ nodes when $\delta = 0.5$, and $E\left[\frac{N}{T}\right] = 0.9675$ with $E[m_T] = 12.35$ when $\delta = 0.8$ (with ARS, we have $E\left[\frac{N}{T}\right] = 0.9962$ with $E[m_T] = 71.60$). Related code is provided at `Matlab-File Exchange` webpage.

## 6 Conclusions

In this work, we have introduced a novel parsimonious ARS scheme (PARS) which automatically reaches a better compromise between acceptance rate and proposal complexity than the standard ARS method. As a consequence, PARS is a faster sampler than ARS.

## 7 Acknowledgements

This work has been supported by the ERC Grant SEDAL, ERC-2014-CoG 647423.

# References

[1] M. Evans and T. Swartz. Random variate generation using concavity properties of transformed densities. *Journal of Computational and Graphical Statistics*, 7(4):514–528, 1998.

[2] W. R. Gilks and P. Wild. Adaptive Rejection Sampling for Gibbs Sampling. *Applied Statistics*, 41(2):337–348, 1992.

[3] Dilan Görür and Yee Whye Teh. Concave convex adaptive rejection sampling. *University College London, Technical Report*, 2009.

[4] W. Hörmann. A rejection technique for sampling from T-concave distributions. *ACM Transactions on Mathematical Software*, 21(2):182–193, 1995.

[5] L. Martino, V. Elvira, and G. Camps-Valls. The Recycling Gibbs Sampler for efficient learning. *arXiv:1611.07056*, 2016.

[6] L. Martino and F. Louzada. Adaptive Rejection Sampling with fixed number of nodes. *arXiv:1509.07985*, 2015.

[7] L. Martino and J. Míguez. A novel rejection sampling scheme for posterior probability distributions. *Proc. of the 34th IEEE ICASSP*, April 2009.

[8] L. Martino and J. Míguez. Generalized rejection sampling schemes and applications in signal processing. *Signal Processing*, 90(11):2981–2995, November 2010.

[9] L. Martino, J. Read, and D. Luengo. Independent doubly adaptive rejection Metropolis sampling within Gibbs sampling. *IEEE Transactions on Signal Processing*, 63(12):3123–3138, 2015.

[10] M. Matthaiou and D.I. Laurenson. Rejection method for generating Nakagami-$m$ independent deviates. *IET Electronics Letters*, 43(25):1474–1475, Dec. 2007.

[11] T. K. Sarkar, Z. Ji, K. Kim, A. Medouri, and M. Salazar-Palma. A survey of various propagation models for mobile communication. *IEEE Antennas and Propagation Mag.*, 45(3):51–82, June 2003.

[12] Q.M. Zhu, X.Y. Dang, D.Z. Xu, and X.M. Chen. Highly efficient rejection method for generating Nakagami-$m$ sequences. *IET Electronics Letters*, 47(19):1100–1101, Sep. 2011.

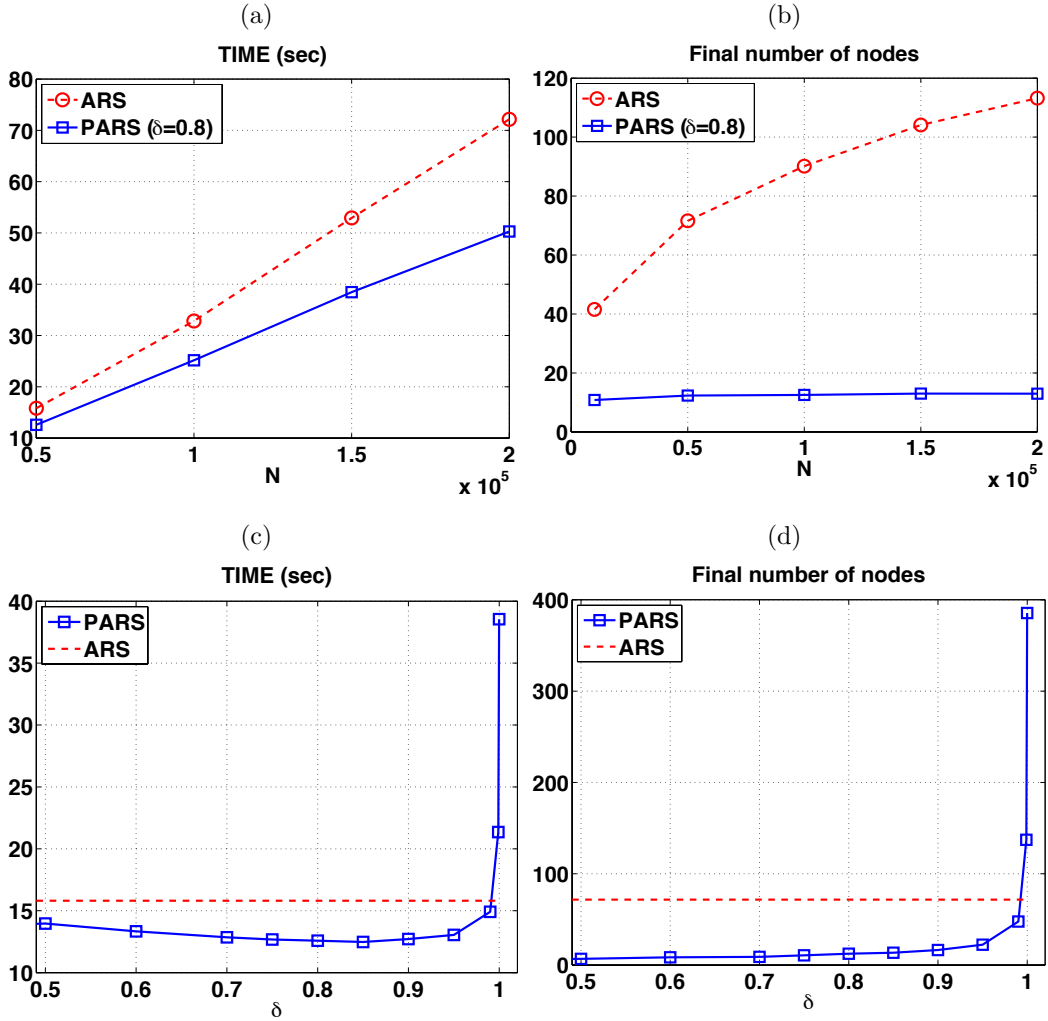

**Figure 2: (a)-(c)** Computational time spent by ARS and PARS as function of the number of the desired samples, $N$, and the parameter $\delta$ (in **(a)** $\delta = 0.8$, in **(c)** $N = 5 \cdot 10^4$). **(b)-(d)** Averaged final number of nodes, $E[m_T]$, as function of $N$ and $\delta$ (in **(b)** $\delta = 0.8$, in **(d)** $N = 5 \cdot 10^4$).